\documentclass[preprint,superscriptaddress]{revtex4-1}
\usepackage{graphicx}
\usepackage{braket}
\usepackage{amssymb,amsmath}
\usepackage[T1]{fontenc}
\usepackage[english]{babel}

\begin{document}

\title{Effects of high pulse intensity and chirp in two-dimensional electronic spectroscopy of an atomic vapor}
\author{M. Binz}
\author{L. Bruder}
\email{lukas.bruder@physik.uni-freiburg.de}
\affiliation{Institute of Physics, University of Freiburg, Hermann-Herder-Str. 3, 79104 Freiburg, Germany}
\author{L. Chen}
\affiliation{Laboratory of Theoretical Physical Chemistry, Institut des Sciences et Ing{\'e}nierie Chimiques, Ecole Polytechnique F{\'e}d{\'e}rale de Lausanne (EPFL), CH-1015 Lausanne, Switzerland}
\author{M. F. Gelin}
\affiliation{School of Science, Hangzhou Dianzi University, Xiasha Higher Education Zone, Building 6, 310018 Hangzhou, China}
\author{W. Domcke}
\affiliation{Department of Chemistry, Technical University of Munich, Lichtenbergstr. 4, 85747 Garching, Germany}
\author{F. Stienkemeier}
\affiliation{Institute of Physics, University of Freiburg, Hermann-Herder-Str. 3, 79104 Freiburg, Germany}

\date{October 19, 2020}

\begin{abstract}
The effects of high pulse intensity and chirp on two-dimensional electronic spectroscopy signals are experimentally investigated in the highly non-perturbative regime using atomic rubidium vapor as clean model system. Data analysis is performed based on higher-order Feynman diagrams and non-perturbative numerical simulations of the system response. It is shown that higher-order contributions may lead to a fundamental change of the static appearance and beating-maps of the 2D spectra and that chirped pulses enhance or suppress distinct higher-order pathways. We further give an estimate of the threshold intensity beyond which the high-intensity effects become visible for the system under consideration.
\end{abstract}

\maketitle

\section{Introduction}
\label{sec:intro}

Coherent two-dimensional (2D) femtosecond (fs) spectroscopy \cite{hamm:2011, jonas:2003, cho:2008, cundiff:2013, kraack:2018} is a powerful nonlinear spectroscopic technique which allows for the study of complicated dynamics and couplings in various quantum systems. 
Usually, 2D spectroscopy experiments are performed with laser intensities in the regime where perturbation theory holds (within the $\chi^{(3)}$ limit) and hence the acquired signals can be described by the third-order response function formalism \cite{mukamel:1995}. However, several new schemes have been developed, for example action-detected variants based on photoionization to study dilute gas-phase samples \cite{bruder:2018,bruder:2019,roeding:2018}, 2D spectroscopy combined with microscopy to reach high spatial resolution \cite{baiz:2014,goetz:2018,tiwari:2018} and higher-order techniques capable of providing information not accessible by third-order schemes \cite{fulmer:2005,bruggemann:2011,dai:2012,zhang:2013,gao:2016,dostal:2018,yu:2019,suss:2019,mueller:2019}. In many of these implementations, higher laser intensities than usual may be reached (e.g. due to tight focusing in spatially resolved experiments), or may be of advantage (e.g. to enhance weak signals from dilute samples) or are even required (e.g. in higher-order spectroscopy schemes).

To the best of our knowledge, there is only limited literature available on the influences of high intensities on the appearance of 2D spectra and possible distortions/artifacts that may be induced by higher-order effects and saturation. In the theoretical study of Br{\"u}ggemann \textit{et al.} \cite{bruggemann:2007}, non-perturbative calculations were performed which reproduce experimental 2D spectra of the Fenna-Matthews-Olson (FMO) photosynthetic antenna complex in the weak-field limit, i.e. where only a fraction of <10\% of the FMO complexes is excited by every pulse. In the same study, they also performed simulations with higher field strengths, corresponding to the case of populating mostly the two-exciton state. They noticed a slightly broader and less structured spectrum, but the overall shape was conserved, from which they concluded that changing the laser intensity has only little influence on the spectra for such systems. In Ref. \cite{vaughan:2007}, 2D degenerate four-wave mixing measurements on atomic rubidium vapor are presented. In this work, three unexpected features that cannot be explained by third-order perturbation theory have been observed, which might be produced by fifth-order or cascaded third-order signals, but no clear conclusion about the origin of these peaks could be given. Furthermore, nonlinear two-phonon and two-photon interband coherences in InSb were investigated with 2D terahertz experiments \cite{Somma:2016}. As the optical interband dipole of InSb is exceptionally large, these experiments have been performed in the highly non-perturbative regime which has manifested itself in the impulsive off-resonant excitation of the interband coherences. 

Very recently, a systematic theoretical study was published exploring the effects of intense laser fields on 2D electronic spectroscopy (2DES) signals using the concept of non-perturbative response functions \cite{chen:2017}. These simulations based on a vibronic model system indicate the occurrence of peak shape modulations and phase shifts, as well as the enhancement of weak features. On the other hand, these results imply that experiments in the high-intensity regime could easily lead to a misinterpretation of 2D spectroscopic data.

Likewise, it is well known from coherent control experiments that chirped laser pulses can have a significant impact on coherent excitation schemes and population transfer \cite{kosloff:1989,assion:1996,prokhorenko:2006,shu:2012,cerullo:1996,nuernberger:2007,conde:2010,simon:2011,debnath:2012,schneider:2011,vitanov:2001,wu:2011}. Yet, the effect of chirped laser pulses on 2DES has not been much explored. Tekavec \textit{et al.} studied peak shape distortion of 2DES for a two-level system interacting with a solvent caused by chirped pulses, both experimentally and via calculations of the third-order response of the system \cite{tekavec:2010}. They introduced a chirp-correction scheme for 2D experiments with a continuum probe \cite{tekavec:2012}. The impact of chirp on peak shapes in 2D 
spectra was studied analytically in Ref. \cite{smallwood:2017}.

In the current work, we systematically investigate, in a combined experimental and theoretical study, the effects of high laser intensities and chirp on 2DES. Our study is based on action-detected 2DES using phase-modulation/phase-cycling, but our results can also be transferred to non-collinear phase-matching based 2DES. As spectroscopic system, we chose atomic rubidium (Rb) vapor which provides us with a clean model system with well separated, narrow absorption features to facilitate the direct observation of peak distortions, amplitude modulation and unexpected features. This is in contrast to previous studies on systems with strongly broadened and overlapping features where subtle effects are much more difficult to identify \cite{bruggemann:2007}. The study of Rb vapors has recently also gained more interest due to the application of coherent 2D spectroscopy to ultracold samples \cite{bruder:2018}. Furthermore, the influence of the intensity of the pump and the probe pulses on transient absorption of atomic Rb vapor beyond the $\chi^{(3)}$ limit has been investigated very recently in Ref. \cite{wang:2019}. 

\section{Experimental setup and method}
\label{sec:exp_setup}

\begin{figure}[b]
\centering\includegraphics[width=0.8\textwidth]{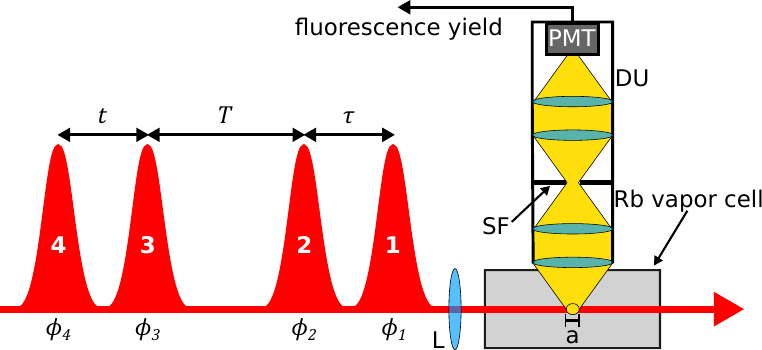}
\caption{Experimental scheme. The Rb atoms in the vapor cell are excited by four phase-modulated pulses 1-4 (see text) with adjustable delays $\tau$, $T$ and $t$. The nonlinear population of the sample is measured via the fluorescence yield. The detection unit (DU) to collect and measure the fluorescence light consists of four lenses, a spatial filter (SF) with minimal diameter $a=0.8\,\mathrm{mm}$ and a photo-multiplier tube (PMT). To acquire spectra in the high-intensity regime, a lens (L) with focal length $f=100\,\mathrm{mm}$ is optionally inserted to focus the laser into the vapor cell.}
\label{fig:exp_scheme}
\end{figure}
A scheme of our experiment is depicted in Fig. \ref{fig:exp_scheme}.
We employ a collinear action-detected type of 2DES for our experiments, that is, we are measuring the nonlinear population of our sample after the interaction with four fs laser pulses. More precisely, we exploit a phase-modulation approach combined with fluorescence detection and lock-in demodulation, as established by Marcus and coworkers \cite{tekavec:2007}. Our experimental setup and measurement scheme are described in more detail elsewhere \cite{bruder:2018,bruder:2019}. Very briefly, the output of a noncollinear optical parametric amplifier (NOPA) with 200\,kHz repetition rate is converted into a collinear four-pulse sequence with adjustable delays $\tau$, $T$ and $t$. The laser pulses resonantly excite room temperature ${}^{87}$Rb vapor contained in a glass cell and its fluorescence is detected with a photo-multiplier tube (PMT) perpendicular to the laser propagation direction.

The carrier-envelope phase (CEP) of each of the four pulses is modulated on a shot-to-shot basis at distinct frequencies via acousto-optic modulators (AOM). Combined with the high repetition rate of the laser, this leads to a quasi continuous modulation of the relative CEPs between the laser pulses in the low kHz regime. Similar to phase-cycling approaches \cite{tan:2008}, the rephasing (RP) and non-rephasing (NRP) signal contributions ($S_{\mathrm{RP}}$ and $S_{\mathrm{NRP}}$) can be extracted from the total signal due to their specific phase signatures. Explicitly, the phase cycling conditions for the RP and the NRP signals are given by:
\begin{align}
\label{eq:phase-cycling_1}
   \phi_\mathrm{RP} = -\phi_1 + \phi_2 + \phi_3 - \phi_4 = \phi_{21} - \phi_{43},\\
\label{eq:phase-cycling_2}
    \phi_\mathrm{NRP} = -\phi_1 + \phi_2 - \phi_3 + \phi_4 = \phi_{21} + \phi_{43}.
\end{align}
Here, it is $\phi_{21} = \phi_2 - \phi_1$ and $\phi_{43} = \phi_4 - \phi_3$. We choose the phase-modulation condition in a way that $\phi_{\mathrm{RP}} = 5\,\mathrm{kHz}$ and $\phi_{\mathrm{NRP}} = 13\,\mathrm{kHz}$. Lock-in demodulation of the total fluorescence yield with respect to these two frequencies extracts then $S_{\mathrm{RP}}(\tau,T,t)$ and $S_{\mathrm{NRP}}(\tau,T,t)$. A Fourier transformation (FFT) of $S_{\mathrm{RP}}(\tau,T,t)$ and $S_{\mathrm{NRP}}(\tau,T,t)$ with respect to the time delays $\tau$ and $t$ yields the complex-valued RP and NRP 2D frequency spectra $\tilde{S}_{\mathrm{RP}}(\omega_{\tau},T,\omega_{t})$ and $\tilde{S}_{\mathrm{NRP}}(\omega_{\tau},T,\omega_{t})$, respectively. The sum of these two spectra gives the 2D frequency-correlation spectrum $\tilde{S}_{\mathrm{C}}(\omega_{\tau},T,\omega_{t})$. Note that our definition of the NRP signal phase-signature is different from the standard definition: $\phi_\mathrm{NRP} = +\phi_1 - \phi_2 + \phi_3 - \phi_4$. We account for the different convention in our data analysis by using accordingly adapted signs in the Fourier transformation of our data to obtain the proper rephasing, non-rephasing and correlation spectra. Throughout this work, we use the convention to plot the excitation frequency ($\omega_{\tau}$) on the x-axis and the detection frequency ($\omega_t$) on the y-axis in the 2D spectra. Data is recorded by scanning $\tau$ and $t$ from 0\,fs to 2960\,fs in 80\,fs steps. Furthermore, we apply zero-padding (factor 2) and multiply the time-data by a Gaussian window prior the Fourier transformation to reduce Fourier transform artifacts.

Measurements in the low-intensity regime are performed with the unfocused laser having a beam diameter ($1/e^2$) of $\approx 2.3 \,\mathrm{mm}$ and a pulse energy of $\approx 33\,\mathrm{nJ}$ per excitation pulse. Acquiring spectra beyond the perturbative regime is achieved by focusing the same laser pulses with a lens ($f=100\,\mathrm{mm}$) down to a focal beam diameter of $\approx 44\,\mathrm{\mu m}$. Furthermore, our pulses are stretched to $\approx 185\,\mathrm{fs}$ which corresponds to an estimated quadratic chirp of $\approx +1510\,\mathrm{fs}^2$. With this parameter, the peak intensity $I_0$ for the low-intensity measurement is about $ 8\,\mathrm{MW}/\mathrm{cm}^{2}$ and that for the high-intensity measurement $\approx 22\,\mathrm{GW}/\mathrm{cm}^{2}$. For comparison, the intensities of the pump and probe pulses used in Ref. \cite{wang:2019} to reveal high-intensity effects in transient absorption signals of Rb vapor were within 10-60\,GW/cm${}^{2}$.

To collect the fluorescence and guide it to the PMT, we use a 4-f lens mapping with a spatial filter (SF) (iris diaphragm, opening diameter $a \geq 0.8\,\mathrm{mm}$) implemented (Fig. \ref{fig:exp_scheme}). The reasons for using a spatial filter are (i) the suppression of stray light and (ii) to restrict the fluorescence detection volume to a small region around the focal spot. Theoretically, for a perfectly aligned detector, the detection volume should have a diameter of $a$ along the beam propagation axis (see Fig. \ref{fig:exp_scheme}). This masking of the fluorescence is important to suppress signals originating from low intensity laser excitation occuring outside the laser focus volume.

\begin{figure}[t]
\centering\includegraphics[width=0.8\textwidth]{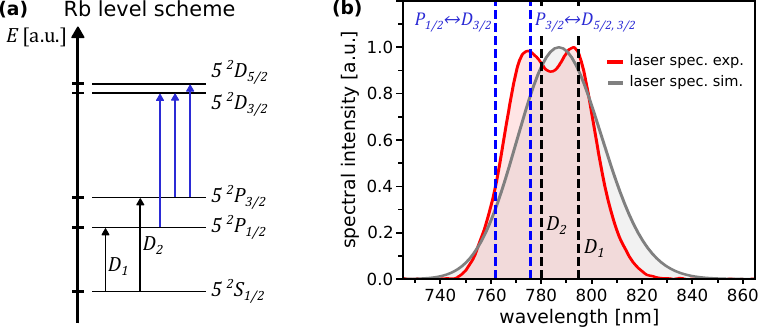}
\caption{Level-scheme, transitions and laser spectrum. (a) Energy level diagram for the relevant transitions in Rb. For the spectral bandwidth of the optical pulses, the system behaves as a five-level system composed of the ground state $5^2S_{1/2}$, two lower-lying electronically excited states $5^2P_{n}$ ($n = 1/2, 3/2$) plus two higher-lying electronically excited states $5^2D_{m}$ ($m = 3/2, 5/2$). Possible ground-excited state transitions ($D$-line transitions named $D_1$ and $D_2$) are indicated by black, excited-state absorption by blue arrows. (b) Experimental laser spectrum (red) compared to the spectrum used as input for simulations (grey), both normalized to one. The wavelengths of the transitions are indicated by the blue and black dashed lines.}
\label{fig:level_scheme}
\end{figure}

We deliberately choose atomic Rb vapor as a target system. This gas-phase system provides well-known sharp spectral lines, while the level structure providing first and second excited state manifolds serves as an adequate model system for common molecular systems. The relevant levels and allowed transitions of Rb atoms along with a typical laser spectrum used in the experiments are shown in Fig. \ref{fig:level_scheme}.
Note that the resolution of the present measurements does not allow to resolve the energy splitting between the two $D_m$ states ($m = 3/2, 5/2$). The transition wavenumbers and transition dipole moments used as input parameters for the numerical simulations of the 2D spectra are listed in Table \ref{tab:transition_dipoles}.
\begin{table}[t]
    \centering
    \begin{tabular}{cccc}
        \hline\hline
       transition & label & $\tilde{\nu}_{ij}\,[\mathrm{cm}^{-1}]$ & $\mu_{ij}\,[ea_0]$ \\
        \hline
        \hline
        $S_{1/2} \leftrightarrow P_{1/2}$ & $D_1$ & 12578.950 & 2.971 \\\hline
        $S_{1/2} \leftrightarrow P_{3/2}$ & $D_2$ & 12816.545 & 4.193 \\\hline
        $P_{1/2} \leftrightarrow D_{3/2}$ & - & 13121.586 & 0.9624 \\\hline
        $P_{3/2} \leftrightarrow D_{3/2}$ & - & 12883.991 & 0.3223 \\\hline
        $P_{3/2} \leftrightarrow D_{5/2}$ & - & 12886.953 & 0.9664 \\\hline
    \end{tabular}
    \caption{Transition wavenumbers and transition dipole moments of the Rb atom. $\tilde{\nu}_{ij}$ is the transition wavenumber (taken from Ref. \cite{nist:2019}) and $\mu_{ij}$ the transition dipole moment in units of $ea_0$, calculated using the transition probabilities listed in Ref. \cite{heavens:1961}.}
    \label{tab:transition_dipoles}
\end{table}

\section{Perturbative description of electronic 2D spectroscopy}
\label{sec:theory}

\begin{figure}[t]
\centering\includegraphics[width=0.9\textwidth]{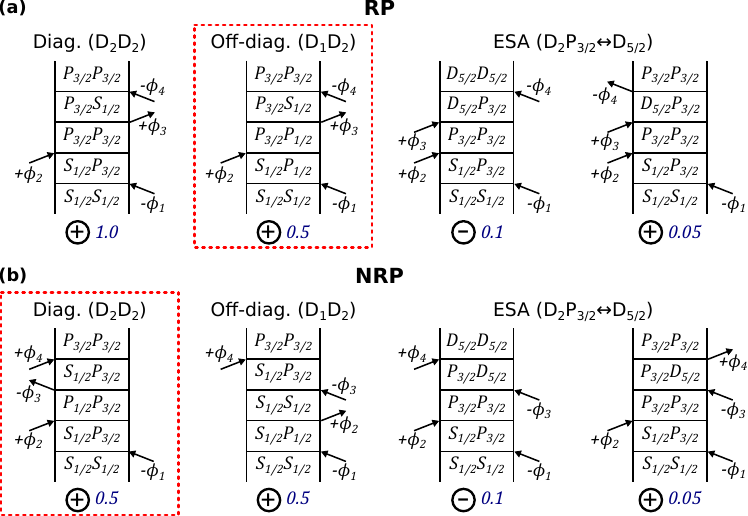}
\caption{4th-order Feynman diagrams. (a) Exemplary diagrams contributing to the rephasing signal. (b) Exemplary diagrams contributing to the non-rephasing signal. The plus and minus signs below the diagrams indicate the sign with which the pathway adds to the overall signal. The factors behind the signs indicate the relative magnitude of the specific contribution (see text). Pathways that beat with respect to $T$ are marked by red dashed boxes.} 
\label{fig:feynman_low}
\end{figure}

A convenient way to keep track of the various signal contributions in 2D spectroscopy are double-sided Feynman diagrams. This diagrammatic description is based on perturbation theory and hence is only valid for adequately weak laser intensities. Nonetheless, it is helpful to discuss the basic features of our signals using Feynman diagrams. In Fig. \ref{fig:feynman_low} we show an exemplary selection of diagrams contributing to the fourth-order signal in collinear population-detected 2D spectroscopy using four excitation pulses and the phase-cycling conditions of Eq. (\ref{eq:phase-cycling_1}) and Eq. (\ref{eq:phase-cycling_2}).

Diagrams contributing to the diagonal peak (where both pump- and probe-frequency are equal to the $D_2$-line transition frequency) are marked ($D_2D_2$). Diagrams contributing to the off-diagonal peak (where the pump-frequency is equal to the $D_1$-line and the probe-frequency to the $D_2$-line frequency) are marked ($D_1D_2$). Excited-state absorption (ESA) diagrams (yielding signal at pump-frequency equals to the $D_2$-line transition and probe-frequency equals to the transition from the $P_{3/2}$ state to the $D_{5/2}$ state) are labeled ($D_2P_{3/2} \leftrightarrow D_{5/2}$).

The following predictions can be deduced from these Feynman diagrams. All ground-state bleach (GSB) and stimulated-emission (SE) pathways, which describe the diagonal and off-diagonal features, carry a positive sign. ESA signals can have either a positive sign (pathways ending in the $P$ state manifold) or a negative sign (pathways ending in the $D$ state manifold). The ESA pathways having a negative sign contribute with a factor of two, as two fluorescence-photons are produced when the atoms end up in a $D_m$ population. Non-radiative de-excitation of excited states can be neglected in the studied dilute sample. Hence, ESA features should appear as negative signals in the 2D spectra. Note that for better convenience we adapted here the sign convention for coherence-detected 2DES and conventional transient absorption pump-probe experiments, i.e. SE/GSB features are positive and ESA negative. To this end, each signal (Feynman diagram) in our population-detected 2DES scheme has been multiplied by an additional factor of -1.

In Fig. \ref{fig:feynman_low}, we also provide a rough estimate of relative amplitudes with which each pathway contributes to the fourth-order population-detected signal. To this end, we assumed a flat laser spectrum and used the dipole moments from Table \ref{tab:transition_dipoles}. All amplitudes are relative to the first RP pathway (labeled $D_2D_2$), which is normalized to 1. Adding up all possible pathways, the strongest ESA feature is roughly a factor of 40-50 weaker than the $D_2D_2$ feature. Hence, the ESA peaks should have negligible amplitudes in the measurements performed in the low-intensity regime.

It is also possible to make a statement about the $T$ dependence of the signal with the help of the Feynman diagrams depicted here. Obviously, pathways that are in a coherent state (red dashed boxes in Fig. \ref{fig:feynman_low}) during the time interval $T$, exhibit a coherent beating with the energy/frequency difference of the excited states $P_{1/2}$ and $P_{3/2}$. This is the case for pathways contributing to off-diagonal spectral features for RP signals, and diagonal spectral features for NRP signals, respectively.

\section{Non-perturbative simulations of the 2D spectra}
\label{sec:sim_methods}

To compare our high-intensity measurements with theory, we perform non-perturbative calculations of 2DES, which has the advantage that all orders of interaction are included automatically and hence the simulated spectra are not restricted to the 4th-order pathways presented above.

Our simulation procedure is similar to that used in Refs. \cite{palacino-gonzalez:2017,palacino-gonzalez:2017a}. It is briefly summarized as follows. The system Hamiltonian $H$  involves five relevant Rb  levels (Fig. \ref{fig:level_scheme}) and is thus described by a $5 \times 5$ diagonal matrix $H_{ij} = E_i \delta_{ij}$, where $i, j =$ $S_{1/2}$,  $P_{1/2}$, $P_{3/2}$, $D_{3/2}$, $D_{5/2}$ and $E_i$ are the corresponding energies. The system-field interaction is treated within the rotating-wave approximation (RWA). The corresponding time-dependent interaction Hamiltonian $H_F(t)$ is also a $5 \times 5$ matrix. It describes 5 dipole-allowed transitions (Fig. \ref{fig:level_scheme}) which are induced by the interaction  of the system with the external fields of four laser pulses involved. The values of the transition frequencies $\tilde{\nu}_{ij}=(E_i-E_j)/\hbar$ and transition dipole moments $\mu_{ij} = \mu_{ji}$ are collected in Table \ref{tab:transition_dipoles}. The spectra of the laser pulses (in RWA) used in the simulations are approximated by the expression  
\begin{equation}
E(\omega) = \exp\left(-\frac{(\omega-\omega_\mathrm{L})^2 t_{\mathrm{FWHM}}^2}{8\ln(2)} - i\frac{\chi_2(\omega-\omega_\mathrm{L})^2}{2}\right),
\label{Ew}
\end{equation}
where $t_{\mathrm{FWHM}}$ is the full width at half maximum (FWHM) pulse duration, while $\chi_2$ is the quadratic chirp and $\omega_{\mathrm{L}} = 2\pi c/(787\,\mathrm{nm})$ denotes the laser central angular frequency (see Fig. \ref{fig:level_scheme} for the comparison of the experimental and simulated $E(\omega)$). The complex pulse envelopes $E(t-t_a)$ in the time domain for each pulse delay $t_a$ ($a =1,2,3,4$) are obtained numerically by the backward Fourier transform of Eq. (\ref{Ew}). The system-field interaction Hamiltonian is then defined as $H_F(t)_{ij}=-\mu_{ij}\sum_a E(t-t_a)$ for $i>j$ and $H_F(t)_{ij}= -\mu_{ij} \sum_a E^*(t-t_a)$ for $i<j$.

For Rb vapor at room temperature, no relaxation/dephasing occurs on the time scale (3\,ps) of the experiment \cite{siddons:2008}. Hence the time-dependent Schr{\"o}dinger equation is appropriate for the description of the photo-induced system dynamics. In matrix-vector notation, it reads
\begin{equation}
i \hbar \partial_t \Psi(t) = (H + H_F(t))\Psi(t). 
\label{Psi}
\end{equation}
In our simulations, Eq. (\ref{Psi}) is solved numerically for all given inter-pulse delay times and phase angles by a fourth-order Runge-Kutta integrator with a time step in the range of 0.5\,fs to 10\,fs depending on the used pulse intensity (see \cite{palacino-gonzalez:2017,palacino-gonzalez:2017a} for the simulation details). The atoms are assumed to be in the ground state before interaction with the laser pulses. Hence the initial condition is
\begin{equation}
\Psi_{S_{1/2}}(\tau_0)=1, \,\,\, \Psi_{n}(\tau_0)=0, \,\,\, n= \, P_{1/2}, \, P_{3/2}, \, D_{3/2}, \, D_{5/2}.
\label{Psi0}
\end{equation}
Here $\tau_0$ is a time moment before the arrival of the first laser pulse (in the simulations, we take $\tau_0 = \tau_1 - 10t_{\mathrm{FWHM}}$). The detected signal is then proportional to the weighted sum of the excited-state asymptotic populations,
\begin{equation}
S(\tau,T,t) \sim |\Psi_{P_{1/2}}(\tau_\infty)|^2 + |\Psi_{P_{3/2}}(\tau_\infty)|^2 +\Gamma|\Psi_{D_{3/2}}(\tau_\infty)|^2 +\Gamma |\Psi_{D_{5/2}}(\tau_\infty)|^2. 
\label{S}
\end{equation}
Here $\tau_\infty$ is a time moment after the arrival  of the fourth laser pulse (in the simulations, we take $\tau_\infty = \tau_4 + 10t_{\mathrm{FWHM}}$), while $0 \le \Gamma \le 2$ is the weighting factor which quantifies contributions of the higher-excited states to the signal (see Refs. \cite {tekavec:2007,lott:2011,maly:2018,kuhn:2020} for an in-depth discussion). In the present case, no non-radiative processes occur and we set $\Gamma = 2$ in all the simulations.

In the present experiment, the RP and NRP contributions $S_{\mathrm{RP}}$ and $S_{\mathrm{NRP}}$ to the signal are discriminated by the phase-modulation procedure \cite{tekavec:2007}. This procedure can be explicitly implemented in perturbative computer simulations of 2DES signals \cite{li:2017,damtie:2017}. The explicit realization  of the phase-modulation in non-perturbative simulations of 2DES signals is, however, numerically inefficient since it requires a long-time propagation of the driven Schr{\"o}dinger equation (\ref{Psi}). Instead, we extract $S_{\mathrm{RP}}$ and $S_{\mathrm{NRP}}$ from the total signal $S$ by using a discrete Fourier transformation (this \cite{wang:2008,gelin:2013b} and similar \cite{renger:2001,mukamel:2004} numerical methods have widely been used in the literature).

The procedure goes as follows. Due to isotropy of space, only relative phases of the pulses matter. Hence all  $\phi_{a}$ can be shifted by e.g. $\phi_{1}$ to yield the modified wave vectors $\bar{\phi}_{a}\equiv\phi_{a}-\phi_{1}$ and the appropriately modified phase-matching conditions (\ref{eq:phase-cycling_1}) and (\ref{eq:phase-cycling_2}). The phase-matched signals $S_{\mathrm{RP}}$ and $S_{\mathrm{NRP}}$ can therefore be computed as follows:
\begin{equation}
S(\tau,T,t)_{\mathrm{RP/NRP}}=\frac{1}{(N_{\phi})^3} \sum_{n_2,\, n_3, \, n_4 = 0}^{N_{\phi}-1} 
e^{-i(\bar{\phi}_{n_2} \pm\bar{\phi}_{n_3} \mp \bar{\phi}_{n_4})}S(\bar{\phi}_{n_2},\bar{\phi}_{n_3},\bar{\phi}_{n_4}).
\label{PNa}
\end{equation}
Here the upper (lower) signs in the exponential function correspond to the RP (NRP) contributions and $\bar{\phi}_{n} \equiv 2\pi n/ N_{\phi}$. Eq. (\ref{PNa}) becomes exact when $N_{\phi} \rightarrow \infty$ and summations are replaced by integrals. If the system-field interaction is not too strong, much smaller values of $ N_{\phi} $ are sufficient. In particular, $ N_{\phi} = 3$ yields the so-called $3 \times 3 \times 3 $ phase cycling scheme, which is known to be exact for weak-field fourth-order signals \cite{tan:2008}. In the context of action-detected 2DES spectroscopy, this procedure has been employed by several experimental groups \cite{tian:2003,de:2014}, while a $5 \times 5 \times 5$ phase cycling scheme was used to extract sixth-order contributions \cite{mueller:2019}. The simulations of the present work are carried out via Eq. (\ref{PNa}) with $ N_{\phi} = 3$. The use of  $ N_{\phi} = 4, 5$ yields virtually indistinguishable signals for the pulse intensities used in our experiments/simulations. In principle, for $N_{\phi} = 3$ double-quantum pathways of the form $2\phi_1 - 2\phi_2 +\phi_3 -\phi_4$ are not cycled out while in the experiment such contributions are discriminated by phase modulation. Yet, for the particular system under study and the given experimental parameters, double-quantum signals are not observed in any of our $N_{\phi} = 3$ simulations and we conclude, that $N_{\phi} = 3$ phase cycling is here sufficient to simulate the experiment.

\begin{figure}[t]
\centering\includegraphics[width=0.3\textwidth]{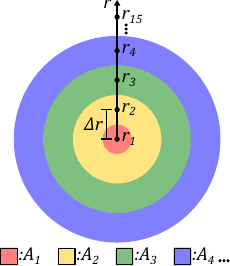}
\caption{Weighting scheme to account for the spatial intensity distribution in the focal point. Here, $r_i$ are the points along the spatial coordinate $r$ for which the intensity is calculated to discretize the intensity distribution and $\Delta_r$ is the distance between these points. The areas $A_i$ represent the weighting factors used to obtain the intensity weighted 2D spectrum from the individual simulations for each intensity (see text).}
\label{fig:weighting_scheme}
\end{figure}

The radial intensity profile in the laser focus follows a Gaussian distribution, which means that the signal arises from a superposition of different laser intensities. To account for this effect, we discretize the expected radial intensity profile at the focus position by calculating the intensity at 15 radial positions $r_1$ to $r_{15}$ (Fig. \ref{fig:weighting_scheme}).
We choose these positions to be equally spaced by $\Delta r$ and range from $r_1 = 0\, \mathrm{\mu m}$ to $r_{15} = 2.1\omega_0$, with $\omega_0$ being the $1/e^2$-radius of the laser beam in the focus. For each of these intensities, a numerical simulation of the 2D signal is performed yielding the RP spectra $\tilde{S}_{\mathrm{RP},i}(\omega_{\tau},T,\omega_{t})$ and NRP spectra $\tilde{S}_{\mathrm{NRP},i}(\omega_{\tau},T,\omega_{t})$ with $i = 1,2,\dots, 15$. These spectra are then added up considering the weighting factors $A_i$ (see Fig. \ref{fig:weighting_scheme}) to obtain the intensity weighted spectra
\begin{equation}
    \label{eq:intens_average_RP}
    \tilde{S}_{\mathrm{RP},\mathrm{weighted}}(\omega_{\tau},T,\omega_{t}) = \sum_{i=1}^{15} A_i \tilde{S}_{\mathrm{RP},i}(\omega_{\tau},T,\omega_{t})
\end{equation}
and
\begin{equation}
    \label{eq:intens_average_NRP}
    \tilde{S}_{\mathrm{NRP},\mathrm{weighted}}(\omega_{\tau},T,\omega_{t}) = \sum_{i=1}^{15} A_i \tilde{S}_{\mathrm{NRP},i}(\omega_{\tau},T,\omega_{t}),
\end{equation}
with
\begin{equation}
    \label{eq:weighting_factor1}
    A_1 = \pi \left(r_1+\frac{\Delta r}{2}\right)^2
\end{equation}
and
\begin{equation}
    \label{eq:weighting_factor2to15}
    A_i = \pi\left(\left(r_i + \frac{\Delta r}{2}\right)^2 -  \left(r_i - \frac{\Delta r}{2}\right)^2\right),
\end{equation}
for $i = 2,3,\dots,15$.

We neglect variations of the intensity along the laser propagation direction (pointing into the drawing plane of Fig. \ref{fig:weighting_scheme}) due to the use of a spatial filter in the fluorescence detection unit (see Sec. \ref{sec:exp_setup}).

\section{Results and discussion}
\label{sec:results}

In the following we present and discuss the effects of high pulse intensity we observed in 2DES experiments. Section \ref{subsec:pertu_vs_nonpertu} deals with the change of the static appearance of 2D spectra when going to the highly non-perturbative intensity regime. It is shown that due to higher-order contributions new peaks can appear and peak shape distortions of the GSB/SE features are possible. In section \ref{subsec:influence_chirp} we show by non-perturbative simulations that chirped pulses have a strong influence on 2D spectra in the high-intensity regime as distinct higher-order pathways can be enhanced or suppressed. We further discuss the effects of high laser intensities on the temporal dynamics reflected in the 2D signals in section \ref{subsec:T_dynamics}. We show that higher-order processes can significantly influence the coherent beat behavior of the 2D maps. Finally, in section \ref{subsec:threshold_intensity}, we present how we define an estimate of the threshold intensity at which high-intensity effects start to play a role for the system under investigation.

\subsection{Perturbative regime vs. highly non-perturbative regime}
\label{subsec:pertu_vs_nonpertu}

\begin{figure}[t!]
\centering\includegraphics[width=0.82\textwidth]{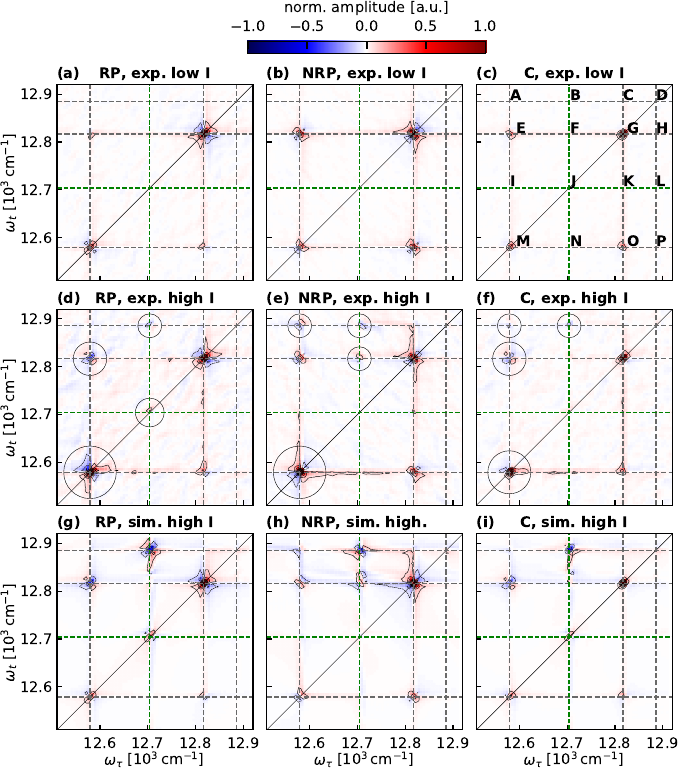}
\caption{Experimental and simulated 2D-spectra: low- vs. high-intensity regime. (a), (b) and (c) show the real parts of the RP, NRP and correlation (C) spectrum for the low-intensity measurement at a population time of $T = 360\,\mathrm{fs}$, (d-f) the respective measurements at high laser intensities and (g-i) the numerical simulation of the high-intensity data assuming laser pulses with a quadratic chirp of $+ 1510\,\mathrm{fs}^2$. All spectra are normalized to a maximal value of 1. Reference lines (dashed) are drawn at the theoretical transition frequencies. The reference line (green dashed) for the $P_{1/2} \leftrightarrow D_{3/2}$ transition is drawn at the expected aliased frequency (see text). The black line indicates the diagonal. Black circles highlight the most prominent differences between low- and high-intensity measurements.}
\label{fig:exp_simu_spec}
\end{figure}
\begin{figure}[b]
\centering\includegraphics[width=0.5\textwidth]{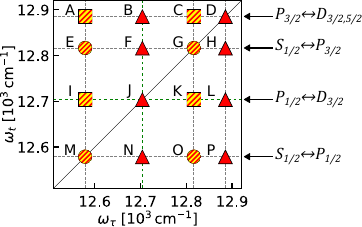}
\caption{Sketch to categorize expected peaks for 4th-order (yellow) and 6th-order (red) processes. The diagram shows the expected peak positions and peak categorization. Peaks are labeled A-P, dashed lines correspond to the expected transition frequencies (see Fig. \ref{fig:exp_simu_spec}). Circles: 4th-order GSB/SE peaks. Squares: 4th-order ESA peaks. Red triangles: peaks allowed only in $\geq$6th-order. Yellow and red shaded symbols: peaks with overlapping 4th-order and $\geq$6th-order contributions.}
\label{fig:categorization}
\end{figure}

In this section, we study the changes observed in our 2D spectra when going from intensities assigned to the perturbative regime to intensities far beyond this regime. For this purpose, Fig. \ref{fig:exp_simu_spec} shows experimental spectra, for a fixed population time of $T=360\,\mathrm{fs}$, acquired using the unfocused laser beam (exp. low I) and data measured by focusing the laser onto the sample (see Sec. \ref{sec:exp_setup}) (exp. high I). The data of the high-intensity experiment are compared with a numerical simulation including the pulse chirp estimated to be present in the experiment (sim. high I). Note that, in the experimental data, saturation of the detector and detection electronics may also lead to the occurrence of intensity-dependent saturation effects stemming from the nonlinear mixing of the linear signal response. The absence of such artifacts was carefully checked. The dashed lines in the spectra indicate the transition frequencies corresponding to all possible transitions in our target system (see Table \ref{tab:transition_dipoles}). Since the frequency resolution of the present measurements is not sufficient to resolve the frequency difference between the two transitions $P_{3/2} \leftrightarrow D_{3/2}$ and $P_{3/2} \leftrightarrow D_{5/2}$, a single reference line is plotted at the arithmetic mean of the two transition frequencies. Furthermore, to speed-up data acquisition and accompanied simulations (see below), large increments (80\,fs) for $\tau$ and $t$ were chosen. As a consequence, the highest frequency features in the spectra ($P_{1/2} \leftrightarrow D_{3/2}$) are aliased and projected onto lower frequencies (green dashed lines).
\begin{table}[t]
    \centering
    \begin{tabular}{ccc|ccc}
        \hline\hline
      label &  pump trans. & probe trans. & label &  pump trans. & probe trans. \\ 
        \hline
        \hline
        A & $S_{1/2} \leftrightarrow P_{1/2}$ & $P_{3/2} \leftrightarrow D_{3/2,5/2}$ & I & $S_{1/2} \leftrightarrow P_{1/2}$ & $P_{1/2} \leftrightarrow D_{3/2}$ \\\hline
        B & $P_{1/2} \leftrightarrow D_{3/2}$ & $P_{3/2} \leftrightarrow D_{3/2,5/2}$ & J & $P_{1/2} \leftrightarrow D_{3/2}$ & $P_{1/2} \leftrightarrow D_{3/2}$ \\\hline
        C & $S_{1/2} \leftrightarrow P_{3/2}$ & $P_{3/2} \leftrightarrow D_{3/2,5/2}$ & K & $S_{1/2} \leftrightarrow P_{3/2}$ & $P_{1/2} \leftrightarrow D_{3/2}$ \\\hline
        D & $P_{3/2} \leftrightarrow D_{3/2,5/2}$ & $P_{3/2} \leftrightarrow D_{3/2,5/2}$ & L & $P_{3/2} \leftrightarrow D_{3/2,5/2}$ &$P_{1/2} \leftrightarrow D_{3/2}$ \\\hline
        E & $S_{1/2} \leftrightarrow P_{1/2}$ & $S_{1/2} \leftrightarrow P_{3/2}$ & M & $S_{1/2} \leftrightarrow P_{1/2}$ & $S_{1/2} \leftrightarrow P_{1/2}$\\\hline
        F & $P_{1/2} \leftrightarrow D_{3/2}$ & $S_{1/2} \leftrightarrow P_{3/2}$ & N & $P_{1/2} \leftrightarrow D_{3/2}$ & $S_{1/2} \leftrightarrow P_{1/2}$ \\\hline
        G & $S_{1/2} \leftrightarrow P_{3/2}$ & $S_{1/2} \leftrightarrow P_{3/2}$ & O & $S_{1/2} \leftrightarrow P_{3/2}$ & $S_{1/2} \leftrightarrow P_{1/2}$ \\\hline
        H & $P_{3/2} \leftrightarrow D_{3/2,5/2}$ & $S_{1/2} \leftrightarrow P_{3/2}$ & P & $P_{3/2} \leftrightarrow D_{3/2,5/2}$ & $S_{1/2} \leftrightarrow P_{1/2}$ \\\hline
    \end{tabular}
    \caption{Assignment of feature labels A, B, C, ..., P (see Figs. \ref{fig:categorization} and \ref{fig:exp_simu_spec}) to the corresponding pump and probe transitions.}
    \label{tab:feature_label}
\end{table}
Table \ref{tab:feature_label} lists all crossing points and hence the positions where signals may appear. We categorize these signals into 4th-order and higher-order signals according to the scheme shown in Fig. \ref{fig:categorization}.

As expected, the low intensity spectra (upper row Fig. \ref{fig:exp_simu_spec}) show only peaks described by 4th-order Feynman diagrams (GSB/SE) with absent ESA contributions due to their much smaller amplitude. To confirm that this measurement can be adequately described by 4th-order perturbation theory (and does not necessarily require non-perturbative treatment), we performed a perturbative simulation (not shown here) using the response functions formalism analogous to the procedure described in \cite{tekavec:2007}. The sign (positive) and the shapes of the features in these simulated 2D spectra match very well with the experimental spectra. In the reminder we use this intensity condition as the reference for 2D spectra in the perturbative regime.

The spectra obtained for high laser intensities (middle row in Fig. \ref{fig:exp_simu_spec}) show a different behavior. The most prominent differences are highlighted by black circles. One consequence of the high intensity is a change in the amplitude ratio of the diagonal peaks G and M. This is visible in the RP, NRP and the correlation spectrum. Furthermore, in the RP spectrum, the off-diagonal feature E changes sign whereas O does not. Hence, an asymmetry is introduced, which is not present in the perturbative regime. For the NRP spectrum this effect is absent. Additionally, new features appear at positions J and B in the RP spectrum and at A, B and F in the NRP spectrum. For other population times $T$ (not shown here) also a feature at position A can be clearly identified in the RP spectrum. The correlation spectrum C combines all the above mentioned changes, as it is the sum of the RP and NRP spectra. However, the additional features, only appearing at high intensities, are less pronounced in the correlation spectrum.

The appearance of the additional peaks when using intense laser pulses can qualitatively be described by Feynman diagrams beyond 4th-order. Such pathways include more than one interaction with a single laser pulse. For example, 6th-order pathways fulfilling the phase-cycling conditions of Eq. (\ref{eq:phase-cycling_1}) and Eq. (\ref{eq:phase-cycling_2}) can be constructed by three interactions with one pulse and one interaction with each of the other pulses in the following way:
\begin{align}
   \phi_\mathrm{RP} = -\phi_1 + \phi_2 + \phi_3 - \phi_4 \pm \phi_j \mp \phi_j, j = 1-4,\\
    \phi_\mathrm{NRP} = -\phi_1 + \phi_2 - \phi_3 + \phi_4 \pm \phi_j \mp \phi_j, j= 1-4.
\end{align}
Hence, these higher order signals are not filtered by phase-cycling and at high intensities may "leak" with significant contributions into the 4th-order signal detection. In analogy, $>$6th-order signals may be constructed that also leak into the 4th-order detection. Due to their nonlinear intensity dependence their amplitudes are expected to be small and are not considered here for simplicity. It is important to note that, although demonstrated here for phase-cycling, the analog case accounts for phase-matching based 2D spectroscopy experiments.

\begin{figure}[t]
\centering\includegraphics[width=0.8\textwidth]{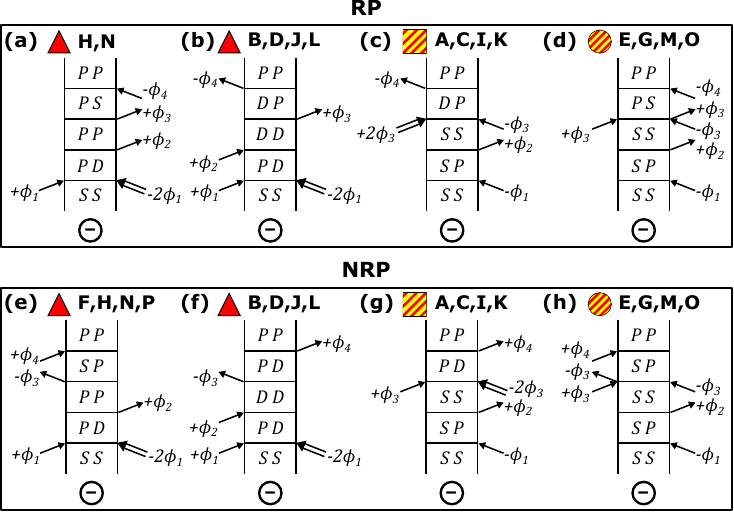}
\caption{6th-order contributions. Typical 6th-order Feynman diagrams fulfilling the phase-cycling conditions for RP pathways (a-d) and NRP (e-h) pathways. The signs below the diagrams indicate the sign with which the pathways contribute to the overall signal. Note that the sign rule for 6th-order pathways is opposite to 4th-order pathways (see text). Similar 6th-order diagrams with positive sign also exist (not shown). For example, pathways similar to (b), (c), (f) or (g), but ending in a $DD$ population, would carry a positive sign. The symbols and labels above the diagrams indicate to which category of features (see Fig. \ref{fig:categorization}) the diagrams can be connected.}
\label{fig:feynman_6th_order}
\end{figure}
Fig. \ref{fig:feynman_6th_order} depicts several representative 6th-order Feynman diagrams to explain the high-intensity spectra. Many more 6th-order pathways exist, but are omitted here for simplicity. Diagrams (a), (b), (e) and (f) can explain the appearance of the new peaks (B, F, J) at positions not allowed by 4th-order pathways (red triangles). Note that no signal is observed at the other positions (D, H, L, N, P) which basically would be allowed by diagrams (a), (b), (e) and (f). This is attributed to a possibly smaller net amplitude of these contributions due to the involved combinations of dipole moments as well as chirp effects (see below). We want to further point out that for peaks F and P, the system has to be in a different $P_n$ state during the time interval $\tau$ than during $t$. This is possible for 6th-order NRP diagrams (b) but not for 6th-order RP diagrams (a) (here only $\geq$8th-order diagrams contribute to F and P). This nicely matches to the fact that the NRP spectrum shows a peak at F and the RP spectrum does not. 

Diagrams (c) and (g) correspond to peak positions (A, C, I, K) where 6th-order contributions can overlap with 4th-order ESA contributions (yellow and red shaded squares). Hence, with the present data, we can not distinguish if the appearance of peak A in the NRP spectrum is caused by 4th-order, 6th-order or by both contributions. Due to the different sign, the overlap of 6th-order contributions (diagrams (d) and (h)) with 4th-order contributions can in prinicple lead to sign and shape changes of the GSB/SE features (E, G, M, O). We observe this effect for peak E in the RP high-intensity spectrum. Using this argumentation, in principle also the off-diagonal peak O should change its sign and shape and the same behavior should be visible in the NRP spectrum (see, e.g. the simulated 2D signals in Ref. \cite{chen:2017}). However, this is not the case for our data. We think that this asymmetry is due to the fact that our laser pulses exhibit a significant chirp. Numerical simulations (see below) show that chirp has a strong influence on the 2D spectra in the high-intensity regime, especially on higher-order pathways. 

While Feynman diagrams offer an intuitive understanding of 2D spectra, they are limited to a perturbative description of nonlinear signals. Therefore, we performed non-perturbative numerical simulations of the high-intensity 2D signals (lower row in Fig. \ref{fig:exp_simu_spec}). We obtain a good agreement between experiment and simulation. The most prominent features are reproduced, although the magnitude and appearance of the high-intensity effects differ to some extent from the experimental data. All additional features corresponding to $>$4th-order pathways are reproduced in the simulations. Also the sign change of peak E is apparent in the calculated RP spectrum. However, the magnitudes of the peaks B, F, J are somewhat overestimated by the simulations. The most striking discrepancy between simulated spectra and experimental spectra is the absence of the strong change in the relative ratio of the diagonal peaks M and G in the simulation.

These differences are attributed to the non-trivial spatial intensity distribution in the laser focus and temporal chirp of the laser pulses (see below). Although we account for these effects in the simulation, it is difficult to match these parameters precisely with the experiment. The chirp was estimated from the measured time-bandwidth product of the pulses without directly measuring the spectral phase of the pulses. The dispersion of the focusing lens and glass walls of the vapor cell were included analytically. The laser intensity can be misgauged if the 4-f lens system is not precisely aligned to the laser focus. Furthermore, the spectral intensity of the laser at the atomic resonances was estimated by assuming a Gaussian spectral profile which deviates to some extent from the experimental laser spectrum (Fig. \ref{fig:level_scheme} (b)).

\begin{figure}[b]
\centering\includegraphics[width=1.0\textwidth]{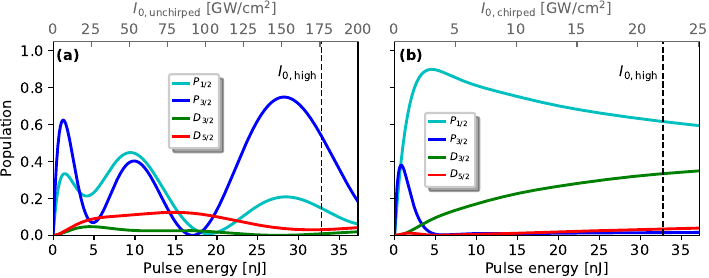}
\caption{Population probability versus pulse energy/peak intensity. Plotted is the numerically calculated population of the states $P_{1/2}$ (cyan), $P_{3/2}$ (blue), $D_{3/2}$ (green) and $D_{5/2}$ (red), after the interaction with a single laser pulse as a function of the pulse energy (bottom scale) and peak intensity (top scale). (a) Assuming a transform-limited pulse. (b) Assuming a pulse exhibiting a quadratic chirp of $+1510\,\mathrm{fs}^2$. Note the lower peak intensities for the chirped and thus temporally stretched pulse. The black dashed lines indicate the pulse energy and corresponding peak intensity ($I_{0,\mathrm{high}}$) used as input for the numerical high-intensity simulations presented in Figs. \ref{fig:exp_simu_spec}, \ref{fig:simu_chirp_vs_unchirped} and \ref{fig:T_dynamics}.}
\label{fig:rabi-osci}
\end{figure}

To set the low and high intensity measurements into perspective of saturation and chirp effects, we have calculated the excitation probability for a single laser pulse exciting the five-level system (5LS) of the Rb atom (Fig. \ref{fig:rabi-osci}).
The high intensity measurements (chirped pulses) are performed at a peak intensity of $\approx 22\,\mathrm{GW}/\mathrm{cm}^2$ (black dashed line in Fig. \ref{fig:rabi-osci} (b)). In contrast, the low intensity measurements (chirped pulses) were performed at a much smaller peak intensity of $\approx 8\,\mathrm{MW}/\mathrm{cm}^2$ (not shown). For a two-level system (2LS) excited by a transform-limited laser pulse, it is well known that the population probability of the excited state, after the interaction with the laser pulse, exhibits an oscillatory behavior as a function of the laser intensity which is the result from Rabi oscillations \cite{allen:1987}. Similarly, in a more complex system, such as the 5LS studied here, the transitions exhibit a clear oscillatory behavior which is, however, changing in frequency and amplitude as a function of the laser intensity (Fig. \ref{fig:rabi-osci} (a)). Panel (b) shows the calculation for the same pulse energies, now including the amount of chirp we estimate to have in the experiments. Obviously, the behavior is completely different for chirped pulses, similar to what was found in Refs. \cite{zhdanovich:2008,kitano:2017}. Here, we observe some strong damping of the excited-state population oscillation by the chirp. Note that Fig. \ref{fig:rabi-osci} shows the population after the interaction with a single laser pulse. For the case of a four-pulse interaction, the dependency on the intensity and the chirp is expected to be more complicated. We therefore expect that errors in the estimates of the experimental parameters chirp and laser intensity are responsible for the mismatch between experiment and simulation.

In this analysis, the Feynman diagrams provided us with an intuitive explanation of the origin of the additional peaks and peak shape distortions appearing at high intensities. In addition, the numerical non-perturbative simulations provide us a quantitative confirmation of these assumptions and show that we can reasonably well simulate the high-intensity effects in the 2D experiments.

\subsection{Influence of chirp in the high-intensity regime}
\label{subsec:influence_chirp}

\begin{figure}[hb]
\centering\includegraphics[width=0.9\textwidth]{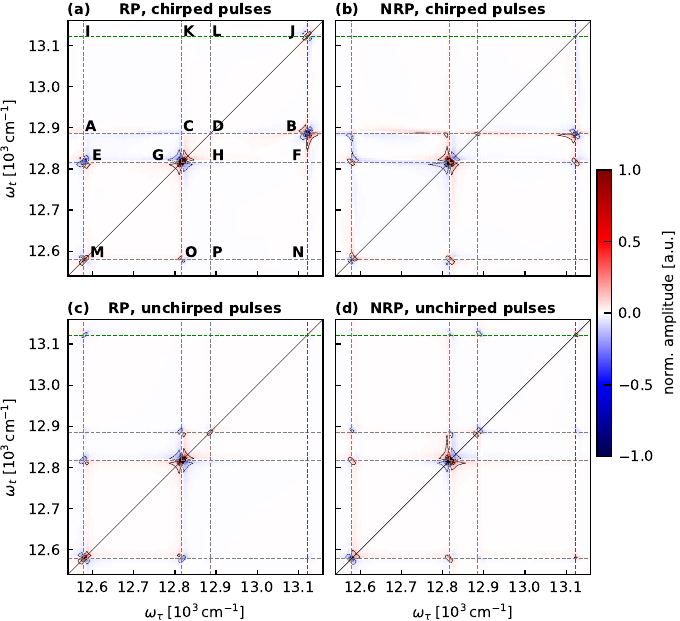}
\caption{Simulated high-intensity spectra showing the influence of chirp. Plotted are numerically simulated RP and NRP spectra (real part, $T = 360\,\mathrm{fs}$) for laser pulses with a quadratic chirp of $+ 1510\,\mathrm{fs}^2$ in panel (a) and (b), as well as for transform-limited pulses in panel (c) and (d). The pulse energy is the same for both simulations. Dashed lines and peak assignments are identical to Fig. \ref{fig:exp_simu_spec}. Note that, in contrast to Fig. \ref{fig:exp_simu_spec}, here fully-sampled spectra are shown (see text) which shifts the green reference line for the $P_{1/2} \leftrightarrow D_{3/2}$ transition from its aliased frequency ($\approx 12700\,\mathrm{cm}^{-1}$) to its fully-sampled frequency ($\approx 13122\,\mathrm{cm}^{-1}$). In all spectra, the most intense peak is normalized to one.}
\label{fig:simu_chirp_vs_unchirped}
\end{figure}

To further investigate the influence of chirp on the 2DES signals, we performed simulations at low and high laser intensities for chirped and transform-limited pulses. At low intensities, we do not observe a significant impact of chirp on the 2D spectra, despite the fairly large amount of the applied chirp (not shown). In contrast, at high laser intensities we observe several differences between the cases of chirped and unchirped pulses (Fig. \ref{fig:simu_chirp_vs_unchirped}). Note that instead of sampling the $\tau$ and $t$ delay with a 80\,fs step size, as done for the data presented in Fig. \ref{fig:exp_simu_spec}, here a step size of 30\,fs is used to avoid aliasing of the $P_{1/2} \leftrightarrow D_{3/2}$ features to lower frequencies (fully-sampled transition freq. in Fig. \ref{fig:simu_chirp_vs_unchirped}: $\approx 13122\,\mathrm{cm}^{-1}$, aliased freq. in Fig. \ref{fig:exp_simu_spec}: $\approx 12700\,\mathrm{cm}^{-1}$, indicated by green dashed lines). Due to the smaller step size, Fig. \ref{fig:simu_chirp_vs_unchirped} shows also a considerably larger frequency range.

Comparing the two simulations it becomes apparent that several spectral features are amplified/damped when introducing chirp (e.g. peaks A, B, C, D, I, J, K, L, N and F). Consequently, it is possible to enhance or suppress distinct higher-order pathways to a certain degree by choosing properly chirped pulses. In combination with the beating analysis (below) our study indicates that the influence of chirp increases for $>$4th-order pathways. This is expected, since all $>$4th-order signals observed with our phase-cycling condition involve multiphoton transitions (multiple interactions with a single laser pulse). For such processes, the strong effect of chirp is in principle well-known from coherent control and population transfer schemes \cite{kosloff:1989,assion:1996,prokhorenko:2006,shu:2012,cerullo:1996,nuernberger:2007,conde:2010,simon:2011,debnath:2012,schneider:2011,vitanov:2001,wu:2011}. As a second observation, the asymmetry in shape, sign and amplitude, induced between the two off-diagonal peaks E and O using intense chirped pulses, is not present when unchirped pulses are employed. Hence, this strong chirp dependence could explain why peaks B, F and J are stronger and the asymmetry between the off-diagonal peaks E and O is more pronounced in the simulated spectra than observed in the experimental data (Fig. \ref{fig:exp_simu_spec}).

\subsection{Investigation of coherent beatings}
\label{subsec:T_dynamics}

\begin{figure}[t]
\centering\includegraphics[width=1.0\textwidth]{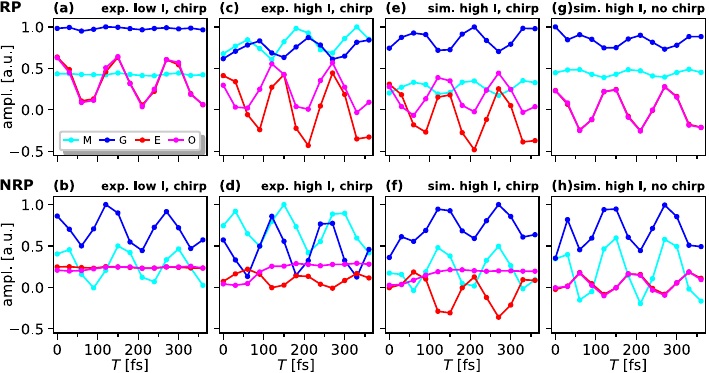}
\caption{Population time dynamics. Plotted are the amplitude of the diagonal (cyan and blue) and off-diagonal (red and magenta) peaks as a function of the population time $T$. The evaluation is done for the real parts of the RP (upper row) and NRP (lower row) spectra. Results for chirped pulses are shown in (a,b) for the low-intensity experiment, in (c,d) for the high-intensity experiment and in (e,f) for high-intensity simulations. (g,h) depicts the dynamics of the simulated high-intensity spectra assuming unchirped pulses. The amplitude is normalized to the respective maximal value in each plot. Note that in (a), (b), (g) and (h) the red and magenta curves are almost perfectly overlapping and hence only the magenta curve is visible.}
\label{fig:T_dynamics}
\end{figure}

Besides the alternation of peak shapes and amplitudes, it is also important to consider the effects of high laser intensities on the temporal dynamics reflected in 2D spectra. To this end, we studied experimentally and theoretically the dependence of the 2D signals on the population time $T$ in the range from 0-360\,fs in 30\,fs steps. Exemplarily, the time evolution of the amplitude of the diagonal (G, M) and off-diagonal (E, O) peaks of the real part of the RP and NRP spectra are shown in Fig. \ref{fig:T_dynamics}. The plotted signal corresponds to the amplitude of the pixel nearest to the theoretical transition frequency in the experimental/simulated 2D spectra.

The low-intensity case follows the predictions from 4th-order Feynman diagrams (Fig. \ref{fig:feynman_low}). An amplitude beating of the off-diagonal peaks is present in the RP contribution, whereas the diagonal peaks exhibit no time dependence. The opposite behavior is true for the NRP spectrum. The frequency of the present beating matches to the predicted value corresponding to the energy difference between the $P_{1/2}$ state and the $P_{3/2}$ state. 

\begin{figure}[t]
\centering\includegraphics[width=0.5\textwidth]{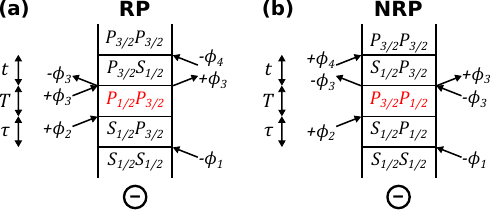}
\caption{6th-order contributions influencing the $T$-dynamics. Exemplary 6th-order diagrams being in a coherent state during $T$ (highlighted red) causing a coherent beating of the diagonal peaks in the RP (a) and the off-diagonal peaks in the NRP (b) spectrum at high intensity.}
\label{fig:6th-order-osci-pathways}
\end{figure}

In the high-intensity regime, the dependence of the signals on the population time changes. Here, also the diagonal peaks show an oscillation with respect to $T$ in the RP spectrum. The time period of this beating is the same as for the off-diagonal peaks (note that the $T$-evolution in both low- and high-intensity regime is governed by the field-free system Hamiltonian), but appears with a $\approx \pi$-phase shift. We attribute this new beating to 6th-order contributions as qualitatively represented by 6th-order Feynman diagrams (Fig. \ref{fig:6th-order-osci-pathways}). Here, RP pathways exist, leading to signal at the diagonal positions M and G, in which the system is in a coherence between the $P_{1/2}$ state and the $P_{3/2}$ state during $T$. A typical diagram for feature G is depicted in Fig. \ref{fig:6th-order-osci-pathways} (a). We further note that the change of the prefactor of the perturbative expansion from $i^4 = 1$ for 4th-order pathways to $i^6 = - 1$ for 6th-order pathways implicates a phase shift of $\pi$ \cite{mueller:2019}. Hence, also the observed phase shift between the diagonal and off-diagonal peak beatings nicely agrees with our assumption of 6th-order contributions overlaying with 4th-order pathways. Note that intensity-dependent phase shifts in the time dynamics have been also detected for transient absorption signals in Ref. \cite{wang:2019}.

For the NRP spectrum, an analogous argumentation applies. Here, 6th-order contributions (e.g. Fig. \ref{fig:6th-order-osci-pathways} (b)) induce oscillatory pathways for the off-diagonal peaks E and O which are not allowed in 4th-order. However, just one of the two off-diagonal features (E) shows the predicted oscillation. The phase of this beat is again roughly the opposite of the features which oscillate due to the 4th-order contributions, at least compared to the diagonal feature G. Our interpretation is confirmed by numerical simulations (Fig. \ref{fig:T_dynamics} (e,f)) where we find exactly the same oscillatory behavior and phase shift.

We want to point out that we observe a non-negligible phase shift between the beatings of the two diagonal peaks (blue and cyan curves) in the low- and high-intensity measurements (Fig \ref{fig:T_dynamics} (b), (c), (d)). We attribute this to be an effect of the data analysis. In the analyzed RP and NRP spectra, peaks comprise of positive and negative parts oscillating out of phase. The phase of the deduced beat thus shows a dependence on the position of the finite-size peak integration area. To reduce this effect, we minimized the integration area to a single pixel of the the 2D maps. On the other hand, increasing the integration area to cover the whole peak would avoid this issue. However, in this case we observe a significant influence on the beat behavior induced by the tails of neighbouring peaks. Our beat analysis thus implies, that even for sparse 2D frequency spectra as analyzed here, precise retrieval of beat phases is difficult. The situation improves for truly absorptive peaks in 2D correlation spectra, yet, flanking ESA features may lead there to a similar phase issue.

Above, we discussed that chirped pulses cause an asymmetry between the off-diagonal peaks E and O in the high-intensity regime. This suggests that the difference in the time dynamics of these two features observed here may be also due to chirp effects. To check this, we repeated the simulation of the $T$-scan with high intensity, but this time assuming transform-limited laser pulses (Fig. \ref{fig:T_dynamics} (g,h)). The pulse energy is kept the same as for the chirped case. As expected, now also a beat of the feature O shows up in the NRP spectrum, having the same frequency, amplitude and phase as the other off-diagonal feature E.

We hence find, that higher-order processes superimposing with 4th-order signals can significantly alter the beat behavior of 2D spectra in amplitude and phase. Chirp can in addition change the beat properties, especially if high-order signal contributions are involved.

\subsection{Intensity threshold for higher-order effects}
\label{subsec:threshold_intensity}

In the present work, we performed 2DES measurements and simulations of a model system at two distinct laser intensities, on the one hand, at a very low intensity in the linear excitation regime and on the other hand, at a much higher intensity where the excitation probability clearly behaves in a nonlinear way and shows Rabi oscillations (Fig. \ref{fig:rabi-osci}). We find good agreement between experiment and simulation. This allows us to use our numerical simulation to explore in more detail the threshold intensity at which distortions due to higher-order effects occur in 2D spectra.

\begin{figure}[b!]
\centering\includegraphics[width=1.0\textwidth]{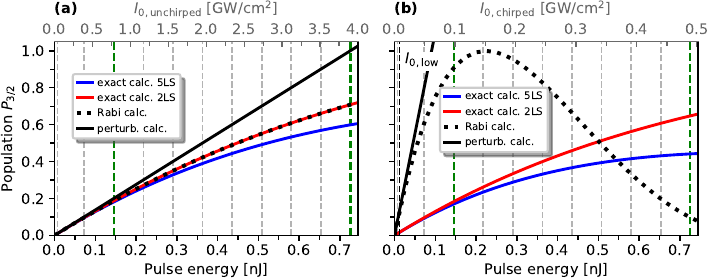}
\caption{Population probability of the $P_{3/2}$ as a function of pulse energy/intensity for cases: perturbative approach (black), Rabi formalism (black dotted), numerically exact solution for a 5LS (blue) and a 2LS (red) assuming a resonant, transform-limited laser pulse (a) and a chirped laser pulse (b). The dashed vertical lines indicate the input intensities used for the simulated 2D spectra. Green dashed lines indicate the parameters used for the 2D spectra in Figs. \ref{fig:onset_dev_5LS} and \ref{fig:onset_dev_2LS}, where the onset of intensity-induced distortions are observed, respectively. The peak intensity ($I_{0,\mathrm{low}}$) of the low intensity 2D measurements discussed further above is also indicated (black dashed).}
\label{fig:popu_perturb_vs_num}
\end{figure}

To this end, we performed numerical simulations of 2D spectra for several laser intensities for a fixed population time of $T = 300\,\mathrm{fs}$. To focus on the transition with the highest transition dipole moment where nonlinearities should be visible first, we shift the carrier frequency from the experimental value ($\approx 787\,\mathrm{nm}$) to the $S_{1/2} \rightarrow P_{3/2}$ resonance (780\,nm). To provide insight how the chirp and the complexity of the target system influence the onset of high-intensity effects, the simulations have been performed for the following cases: (i) the complete five-level system (5LS) excited by transform-limited pulses, (ii) the 5LS excited by pulses with a quadratic chirp of +1510\,fs${}^2$, (iii) just a two-level system (2LS) consisting of ground state $S_{1/2}$ and excited state $P_{3/2}$ excited by unchirped pulses, (iv) the 2LS excited by chirped pulses. We have chosen the input intensities for this simulation series such that the population probability of the $P_{3/2}$ state, in a pure 2LS, interacting with a single transform limited pulse, exhibits the following values: 1\% and 10-100\% in 10\% increments (Fig. \ref{fig:popu_perturb_vs_num} (a)). These values are calculated with the following formula for the excited-state population ($p_e$)
\begin{equation}
    \label{eq:perturb_popu_calc}
    p_e = \frac{|\mu_{eg}|^2\pi I_0 t_{\mathrm{FWHM}}^2}{4 \ln{2} \hbar^2 \epsilon_0 c},
\end{equation}
which is derived from first order perturbation theory adopting the rotating-wave approximation (RWA) and assuming a resonant laser pulse described by
\begin{equation}
    E(t) = E_0 e^{{-2\ln{2}\left(\frac{t}{t_{\mathrm{FWHM}}}\right)^2}}\cos{\left(\omega_{eg}t\right)}.
\end{equation}
Here, $\mu_{eg}$ is the transition dipole moment from the ground ($S_{1/2}$) to the excited state ($P_{3/2}$), $I_0$ the peak intensity of the laser pulse, $t_{\mathrm{FWHM}}$ the full width at half maximum (FWHM) pulse duration with respect to the intensity, $\omega_{eg}$ the pulse carrier frequency and $E_0$ the electric field amplitude. For higher intensities/populations, this simple estimation does not accurately describe the population, as saturation of the transition is not considered (see the numerical exact calculation in Fig. \ref{fig:popu_perturb_vs_num}). Corresponding pulse energies are calculated based on a beam diameter of $44\,\mu\textrm{m}$. The simulations for the chirped pulses are performed for the same pulse energies. 

Fig. \ref{fig:popu_perturb_vs_num} shows the numerically exact calculation of the excitation probability for the simulation cases (i)-(iv) compared to Eq. (\ref{eq:perturb_popu_calc}) and the population calculated by the Rabi formalism as done in Refs. \cite{zhdanovich:2011a, lee:2015}.
For resonant excitation with an unchirped laser pulse, the Rabi formula representing the excited-state population can be expressed as
\begin{equation}
\label{eq:rabi_formular}
    p_e = \sin^2{\left(\frac{\mu_{eg}\sqrt{\frac{2I_0}{\epsilon_0 c}}}{2\hbar}\sqrt{\frac{\pi t^2_{\mathrm{FWHM}}}{2 \ln{2}}}\right)}.
\end{equation}
One can see from Fig. \ref{fig:popu_perturb_vs_num} that the intensities we have chosen for this simulation series (dashed vertical lines) range from the regime where perturbation theory is accurate for the estimation of the population probability after one pulse to the regime where it completely fails. In panel (b), the calculations are plotted for the chirped pulse scenario. The same equations as above have been used for the Rabi and the perturbative calculation (Eq. (\ref{eq:rabi_formular}) and  (\ref{eq:perturb_popu_calc})), but this time plugging in the pulse duration and peak intensity of the chirped pulse. Obviously, these simple formulas fail to estimate the population for strongly chirped pulses, even for very low intensities. 
Fig. \ref{fig:popu_perturb_vs_num} reveals the intensity regime where perturbation theory starts to deviate from the exact numerical simulation, which implies at which intensity the experiment cannot be described by low-order perturbation theory anymore. Note that measuring the fluorescence saturation curve would be an experimental way to estimate this onset regime. We further note, that the values calculated here are valid for the specific system studied here. Yet, our simulations may provide a rough general estimate for the threshold intensity below which perturbation theory is valid. This conclusion is also corroborated by the observation that the breakdown of the perturbative description of transient absorption of pump-probe signals in the displaced harmonic oscillator model starts at $\mu_{eg}\sqrt{2 I_0/(\epsilon_0 c)} > 0.01$ eV \cite{gelin:2013b}. For the transition dipole moment of the $S_{1/2} \rightarrow P_{3/2}$ transition, this yields the threshold field intensity of 0.3 GW/cm$^2$, in good agreement with Fig. \ref{fig:popu_perturb_vs_num}. Note also the different behavior for a 2LS and a 5LS at higher intensities, implying that for more complex systems (e.g. molecular systems) the situation may change. However, at low intensities, both systems behave very similar, suggesting that our conclusions may be generalized to more complex systems. We also point out that we consider here population probabilities (two field-matter interactions exciting a population), which corresponds to the linear absorption of the system, a quantity that is readily experimentally accessible. In 2D spectroscopy it is also common to think in terms of single field-matter interactions. For a 2LS this probability corresponds to the square-root of the population probability.

\begin{figure}[t]
\centering\includegraphics[width=1.0\textwidth]{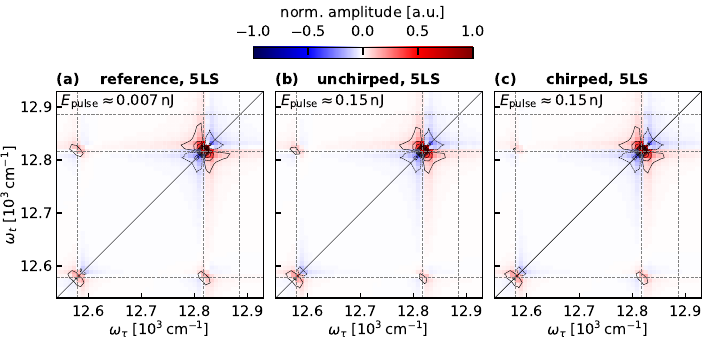}
\caption{Simulated 2D spectra for the Rb 5LS for two different pulse energies. Plotted is the normalized real part of the NRP contribution. Just a zoom onto the relevant part of the spectra is shown. (a) low intensity reference spectrum, (b) simulated spectrum assuming unchirped pulses with a pulse energy of $\approx 0.15\,\mathrm{nJ}$, (c) same pulse energy as in (b), but assuming chirped pulses.}
\label{fig:onset_dev_5LS}
\end{figure}

While Fig. \ref{fig:popu_perturb_vs_num} considers the interaction with a single laser pulse (i.e. two field-matter interactions), for 2DES experiments it is more relevant at which intensities deviations are observed in the nonlinear system response generated by four field-matter interactions. To define the threshold intensity at which higher-order effects appear, we investigated the distortions in the corresponding 2D spectra. To do so, we normalized, for all the simulated cases (i)-(iv), the real parts of the different contributions (RP, NRP, C) for each intensity to have the maximum amplitude of one. The spectra simulated for the intensity corresponding to a population probability of 1\% after one pulse is chosen as reference spectra for the perturbative regime. Our simulations show that distortions are best visible in NRP spectra, for which the onset of distortions are shown in Fig. \ref{fig:onset_dev_5LS} for the simulation cases (i) and (ii) for the 5LS.

For the simple model system of Rb atoms yielding clean, well-separated and sharp spectral features, distortions due to high pulse energies can be directly identified by a qualitative comparison of the spectra with the low-intensity reference spectrum. Here, we find that first visible changes appear for the off-diagonal features. For the given pulse energy in Fig. \ref{fig:onset_dev_5LS} ($\approx 0.15\,\mathrm{nJ}$), these changes are more pronounced for the case of chirped pulses, although the population probability (numerically exact calculation) of the strongest transition induced by one pulse is nearly the same for both cases ($\approx 18\%$ assuming unchirped pulses and $\approx 17\%$ assuming chirped pulses). For population probabilities of 10\%, no distortions can be identified. Hence, we conclude that the limit of the perturbative intensity regime lies in the range of 10-20\% excitation probability of the $P_{3/2}$ state, corresponding to an intensity of $\approx 0.4-0.8\,\mathrm{GW}/\mathrm{cm}^2$ for transform-limited pulses. Note, that the overall signal scaling (integrated absolute spectra) deviates already at lower intensities from the $I^2$ dependency expected for 4th-order processes. However, we did not analyze this in detail as our focus was more on distortions of the spectral appearance.

\begin{figure}[t]
\centering\includegraphics[width=1.0\textwidth]{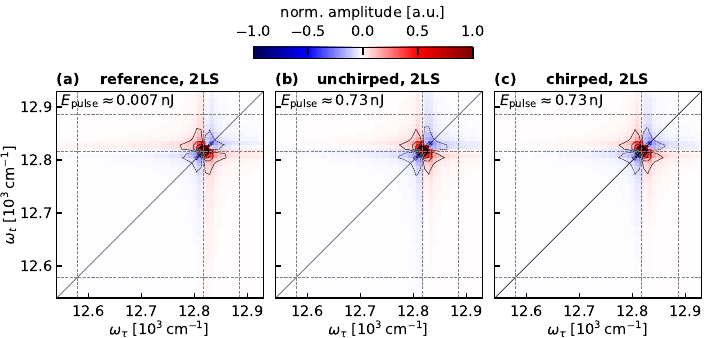}
\caption{Simulated 2D spectra for the Rb 2LS ($S_{1/2}$ ground state and $P_{3/2}$ excited state) for two different pulse energies. Plotted is the normalized real part of the NRP contribution. (a) low intensity reference spectrum, (b) simulated spectrum assuming unchirped pulses with a pulse energy of $\approx 0.73\,\mathrm{nJ}$, (c) same pulse energy as in (b), but assuming chirped pulses.}
\label{fig:onset_dev_2LS}
\end{figure}
For comparison, Fig. \ref{fig:onset_dev_2LS} shows corresponding simulations for a 2LS (cases (iii) and (iv)). Here, we find that even for the highest tested pulse energy ($\approx 0.73\,\mathrm{nJ}$) in this study, which delivers about 71\% excited-state population for transform-limited pulses and 65\% for chirped pulses (numerical exact calculations), the feature in the 2D spectrum does not change much. Hence, for a pure 2LS, distortions due to high laser intensities and chirp seem to be absent and we conclude that distortions arise from the mixing of multiple states during the nonlinear multi-pulse excitation in 2DES experiments.

\begin{figure}[htb]
\centering\includegraphics[width=1.0\textwidth]{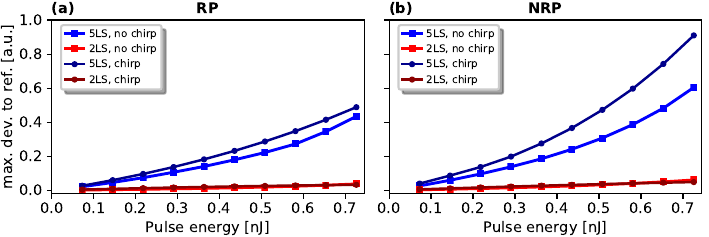}
\caption{Maximum deviation from the reference spectrum vs. pulse energy. The absolute maximum of the difference of the normalized low intensity reference spectra and the spectra at the higher intensities for cases (i)-(iv) are plotted (see text). (a) shows the evaluation for the real part of the RP contribution and (b) for the real part of the NRP contribution. The blue squares depict the results for the 5LS assuming TFL pulses, the red squares for the 2LS assuming TFL pulses, the dark blue points for the 5LS assuming chirped pulses and the dark red points for the 2LS assuming chirped pulses.}
\label{fig:plot_max_dev}
\end{figure}

For a more quantitative analysis we calculated the deviation of the 2D spectra by subtracting the reference spectrum (normalized to the maximum absolute amplitude of one) from the higher intensity spectra, normalized in the same way. The maximum absolute value of the difference spectrum as a function of pulse energy for the cases (i)-(iv) is shown in Fig. \ref{fig:plot_max_dev}.
As already indicated in our qualitative analysis, the deviation in the NRP spectrum increases more strongly than in the RP spectrum and chirping the pulses results in an increased maximal deviation for the 5LS. Furthermore, the maximal difference to the reference spectrum assuming a 2LS does not show such a drastic increase with intensity.

In conclusion, our analysis implies that observable intensity-induced distortions of the 2D spectra start to appear at excitation probabilities of 10-20\% for a single pulse. This is consistent with other work and may be generalized to more complex systems as a rule of thumb. We note, that at a population probability of 20\%, the linear absorption of the sample hardly deviates from a linear curve ($\approx 10\%$ deviation for the case of no chirp) (see Fig. \ref{fig:popu_perturb_vs_num} (a)). Such small deviations are usually difficult to determine in the experiment. Therefore, care has to be taken in determining the critical intensity in each experiment.

\section{Conclusions}

In this work, we investigated effects caused by high intensities and chirped laser pulses on 2D electronic spectroscopy. To this end, we performed collinear phase-modulated 2DES experiments of Rb atoms, which serve as a well-defined model system with well-known and sharp features, in different intensity regimes and compared the results with non-perturbative numerical simulations. The advantage of having sharp and well-separated spectral features in the atomic gas-phase system facilitated the observation of subtle changes in the spectra, such as amplitude modulations and additional peaks appearing at high intensity. On the other hand, the absence of dephasing on the observation time scale (3\,ps) required the application of a window function (here Gaussian) to the data to avoid Fourier transform artifacts \cite{hamm:2011}, resulting in artificial lineshapes with no physical meaning. Hence, the investigation of lineshape effects \cite{tekavec:2010} was not possible with our data. 

We found that even though we use a high-repetition rate laser featuring low pulse energies ($\approx 30\,\mathrm{nJ}$), high intensities in the regime of multiple Rabi-cycles are already reached by focusing the laser onto the target system with a moderately short focal length of $f = 100\,\mathrm{mm}$ (Fig. \ref{fig:rabi-osci}). As a consequence, new features appear in the 2D spectra which can be explained by 6th-order contributions that cannot be discriminated by phase-modulation/cycling or phase-matching. Moreover, higher-order contributions also influence the peak shape, sign and amplitude of the features associated with GSB/SE pathways (diagonal and off-diagonal peaks). We further found that also the beating behavior of these peaks changes due to superimposing higher-order signals. The different beating behavior for diagonal and off-diagonal features in RP and NRP 2D spectra, respectively is commonly used to separate coherent from incoherent dynamics \cite{cheng:2008}. We find, that at high laser intensities the beat behavior of RP and NRP 2D spectra intermix, which in principle compromises such analysis. Besides the change in the relative ratio of the diagonal GSB/SE features, all the experimentally observed effects are qualitatively reproduced by non-perturbative simulations. Although shown here for phase-cycling, analogous effects are expected for phase-matching experiments.

To investigate the influence of chirp in the high-intensity regime, we compared simulations for chirped (+1510\,fs${}^2$) and transform-limited pulses. Here, we found that it is possible to enhance or suppress distinct higher-order pathways using chirped pulses. Furthermore, an asymmetry between the off-diagonal peaks, as observed in the experimental data, is present when using chirped pulses, but vanishes for transform-limited pulses.

As we observe all these effects already at rather low pulse energies and moderate focusing conditions, this raises the question at which intensity regime the onset of high-intensity effects occurs in 2D spectra. To this end, we performed simulations for several laser intensities. We found that the onset of high-intensity effects is best visible in the NRP contribution and manifests itself primarily in changes of the off-diagonal features. For a given pulse energy, the changes are more pronounced in the case of chirped pulses. For transform-limited pulses, we conclude that the high-intensity limit lies in the range of $\approx 0.4-0.8\,\mathrm{GW}/\mathrm{cm}^2$, corresponding to an excitation probability per pulse of 10-20\% of the $P_{3/2}$ state (transition with the highest transition dipole moment). While these estimates are valid for a specific system (Rb atom as five-level system), they may be used as a general qualitative estimate of the threshold intensity below which the spectral appearance of 2D spectra can be described by perturbation theory. 

\begin{acknowledgments}
M.B., L.B. and F.S. acknowledge funding by the European Research Council (ERC) with the Advanced Grant COCONIS (694965) and by the Deutsche Forschungsgemeinschaft (DFG) IRTG \textit{CoCo} (2079). M.F.G. acknowledges support of Hangzhou Dianzi University through the startup funding. L.P.C. acknowledges the financial support from the European Research Council (ERC) under the European Union's Horizon 2020 research and innovation programme (Grant Agreement No. 683069-MOLEQULE).
\end{acknowledgments}

\bibliography{high_intens_2D_arxiv_updated}

\end{document}